\documentclass[]{aastex631}

\usepackage{xspace}
\usepackage{graphicx}
\usepackage{gensymb}

\begin{document}
\def\hess{H.E.S.S.\xspace}
\def\fermi{\textit{Fermi}-LAT\xspace}
\def\ssftt{SS~433\xspace}
\def\grs{GRS~1915+105\xspace}
\def\xray{X-ray\xspace}
\def\ours{PS~J1915.5+1056\xspace}

\title{Persistent GeV counterpart to the microquasar \grs}

\author[0000-0003-0766-6473]{Guillem Martí-Devesa}
\affiliation{Dipartimento di Fisica, Universit\'a di Trieste, I-34127 Trieste, Italy}

\affiliation{Istituto Nazionale di Fisica Nucleare, Sezione di Trieste, 34127 Trieste, Italy}

\author[0000-0002-9105-0518]{Laura Olivera-Nieto}
\affiliation{Max-Planck-Institut f\"{u}r Kernphysik, P.O. Box 103980, D 69029, Heidelberg, Germany}

\correspondingauthor{Guillem Martí-Devesa, Laura Olivera-Nieto}
\email{guillem.marti-devesa@ts.infn.it, laura.olivera-nieto@mpi-hd.mpg.de}

\begin{abstract}
Microquasars are compact binary systems hosting collimated relativistic jets. They have long been proposed as cosmic-ray accelerators, probed via the gamma-ray emission produced by relativistic particles. However, the observational evidence is steadily increasing but limited: there are around twenty microquasars known to date, of which only three have so far been firmly detected in the GeV gamma-ray range, always in a flaring or special spectral state. Here we present \fermi observations of the region around the microquasar \grs, which reveal the presence of previously unknown multi-GeV emission consistent with the position of the microquasar. No periodicity or variability is found, indicating a persistent source of gamma rays. The properties of the emission are consistent with a scenario in which protons accelerated in the jets interact with nearby gas and produce gamma rays. We find that if the jet has been operating at an average of 1\% of the Eddington limit for 10\% of the time that \grs spent in its current mass-transfer state, the transfer of 10\% of the available power to protons is enough to reach the $\sim 3 \cdot 10^{49}$ erg required to explain the GeV signal. Therefore our results support a scenario in which microquasars with low-mass stellar companions act as hadronic accelerators, strengthening the idea that microquasars as a class contribute to at least some fraction of the observed cosmic-ray flux.
\end{abstract}


\section{Introduction} \label{sec:intro}
Stellar binary systems hosting a compact object reveal themselves through the \xray emission associated with accretion onto the compact object, earning them the name of \xray binaries. Microquasars are a subclass of \xray binaries in which a collimated relativistic outflow of plasma (jet) is formed as a consequence of the aforementioned accretion~\citep{Mirabel1999, Fender2004a}.

The microquasar \grs was first detected as an \xray source by the WATCH instrument on board of the GRANAT observatory in 1992~\citep{CastroTirado1992}. Follow-up observations with the VLA~\citep{Mirabel1994,Rodriguez1999} and MERLIN~\citep{Fender1999} in the radio band revealed a highly variable counterpart with apparently superluminal two-sided ejections. This was the first observation of relativistic motions in an object inside our Galaxy, and implied intrinsic velocities for the ejecta near the speed of light. These findings established the presence of a jet with velocity $v\sim0.8$c and  angle to the line of sight $\theta\sim63$\degree~\citep{Fender1999,Rodriguez1999}, making the system a microquasar. 

After some initial constraints on the distance by \cite{Fender1999} and \cite{Chapuis2004}, recent parallax measurements result in a distance estimate of $d =9.4\pm0.6\pm0.8$~kpc~\citep{Reid2023}, which we will adopt through this work.

The mass of the black hole has been subject to debate, with an initial claim of  M$_{\mathrm{BH}}=14$~M$_{\odot}$~\citep{Greiner2001a} which was later revised to lower values ranging between 10~M$_{\odot}$~\citep{Steeghs2013} and 12~M$_{\odot}$~\citep{Reid2014,Reid2023}. The corresponding Eddington luminosity is on the order of L$_{\mathrm{Edd}} = 1.26\cdot 10^{38} \cdot \mathrm{M}_{\mathrm{BH}} $~erg~s$^{-1}M_\odot^{-1}\sim 10^{39}$~erg~s$^{-1}$. The spectral type of the donor star was identified to be a K-type star~\citep{Greiner2001} with a mass of 0.5$M_\odot$\citep{Steeghs2013}, making \grs a low-mass \xray binary (LMXB). Its orbital period is $33.85\pm0.16$~days~\citep{Steeghs2013}.

Since its detection more than 30 years ago, \grs has been extensively monitored from radio to \xray energies, revealing a highly variable source with distinct behaviours across wavelengths. The emission in the radio band is characterised by quasi-periodic oscillations~\citep{Pooley1997,Fender2000}, seen also in the infrared band~\citep{Fender1998a}. The source flares episodically in both radio and X-rays, with the same anti-correlation between these two bands that is seen for other microquasars. Most notably, \grs had been a consistently bright \xray source since its discovery, implying intrinsic \xray luminosities in the range of 0.1 to 1 L$_{\mathrm{Edd}}$, until a sudden decrease in June 2018. An initial decrease brought the flux down by an order of magnitude, followed by a continuous decline of the \xray emission while flaring activity increased at other wavelengths. This behaviour has been attributed to either an obscuration~\citep{Motta2021, Balakrishnan2021} or an intrinsic change into a hard state~\citep{Fender2004a,Koljonen2021}.

Gamma-ray emission from microquasars has long been predicted~\citep[e.g.][]{Atoyan1999, Aharonian2004, valenti}, both in the high-energy (HE, GeV-scale) and very-high energy (VHE, TeV-scale) bands of the spectrum, usually invoking particle acceleration inside the jets. Only two objects, both high-mass \xray binaries (HMXB), have so far been firmly detected as GeV sources: Cyg~X-3~\citep{Tavani2009,Collaboration2009,Corbel2012}, where orbital gamma-ray modulation is seen and Cyg~X-1~\citep{Zanin2016, Zdziarski2017}, where gamma-ray emission is detected during its hard \xray state. Additionally, a source has been detected near SS~433~\citep{Li2020}, showing flux variability correlated with the jet precession period. At higher energies only two microquasars have been firmly detected in the VHE band, both as persistent sources: SS~433~\citep{Abeysekara2018,HESSCollaboration2024,Cao2024} and the intermediate-mass \xray binary (IMXB) V4641~Sgr~\citep{V4641sgr}. No significant VHE emission has been reported from any of the other HE sources~\citep{Ahnen2017,Albert2021}. 

In comparison, the gamma-ray detection of LMXBs remains elusive despite intensive searches~\citep[particularly during the \xray flaring states of V404~Cygni;][]{Loh2016,Piano2017, Xing2021, Harvey2021}. Previous searches for HE emission from \grs placed only flux upper limits~\citep{Olaf03,Bodaghee2013}. Intriguingly, excess emission was reported by HEGRA at VHE~\citep{Aharonian1998}, a hint not confirmed by follow-up observations with either the H.E.S.S.~\citep{HESSCollaboration2018} or MAGIC~\citep{Saito2009} telescopes.

The relative proximity to Earth of microquasars compared to that of jetted AGNs makes them a critical tool in the study of particle acceleration in jets. However, as every system detected so far displays a unique phenomenology in the gamma-ray band, the role of microquasars as a population remains ambiguous, especially for LMXBs. The recent detection of two systems in the VHE band suggests that microquasars could provide intermittent contributions to the Galactic cosmic-ray sea
at energies up to a few PeV~\citep{HESSCollaboration2024}. However, setting constraints on the magnitude of this contribution remains challenging given the low number of systems detected. Studying the gamma-ray emission from a diverse sample of objects (covering a wide range of jet power, velocity, orientation, and companion mass), is crucial to disentangle the phenomena specific to individual sources from universal properties of particle acceleration in fast-moving jets within our Galaxy.

 In this letter we present evidence for a GeV counterpart to the microquasar \grs using the latest \fermi data, including almost four times as much data as in the latest study dedicated to the same source~\citep{Bodaghee2013}. The paper is structured as follows: Section~\ref{sec:analysis} reports on the observation and analysis details; Section~\ref{sec:results} presents the results of the spectral, morphological, and flux variability analyses; Section~\ref{sec:discussion} discusses the possible counterparts and evaluates the feasibility of the association with \grs; and Section~\ref{sec:conclusion} summarises our conclusions.

\section{Observations and analysis} \label{sec:analysis}
We use data from the Large Area Telescope (LAT), a pair-production gamma-ray detector onboard the \textit{Fermi Gamma-ray Space Telescope} \citep{LATpaper}. The LAT is in operation since August 2008 and thanks to its wide field of view, it surveys the whole sky in photon energies from 30 MeV to more than 100 GeV every 3~h. The LAT's angular resolution depends strongly on the photon energy, with the point-spread function (PSF) 68\% containment radius ranging between several degrees
at the lowest energies to less than $0.1^{\circ}$ above 10 GeV \citep{4FGL}. To analyse the LAT data we use the \texttt{Fermitools} \citep[v2.2.0;][]{Fermitools} and \texttt{fermipy} \citep[v1.2.0;][]{Wood17} packages.

We select almost 16 years (from August 4 2008 to May 22 2024) of P8R3 \texttt{SOURCE} data \citep[evclass $=128$;][]{Atwood13, Bruel18} from 1 to 100 GeV within $10^\circ$ of \grs. The motivation for the 1~GeV threshold is two-fold: to minimise the impact of the bright diffuse emission in our analysis and take advantage of the improved PSF at higher energies. Additionally, we apply a cut on the zenith angle $z_{\rm max}$ at $105^{\circ}$ to prevent Earth-limb contamination. The selected gamma-ray events are binned into a three-dimensional data cube, consisting of a sky-map of dimensions $10^{\circ}\times10^{\circ}$ (our region of interest, ROI) and $0.04^{\circ}$ bin size, centred at the position of \grs and an energy axis with $16$ energy bins equally spaced in logarithmic energy between 1 and 100 GeV. 

We fit combined spatial and spectral models to this data using a binned maximum-likelihood framework~\citep{Mattox96}. The analysis is done jointly for the PSF0 to PSF3 event types\footnote{\url{https://fermi.gsfc.nasa.gov/ssc/data/analysis/documentation/Cicerone/Cicerone_Data/LAT_DP.html}}. We use the gll\_iem\_v07.fits and iso\_P8R3\_SOURCE\_V3\_PSF3\_v1.txt as the Galactic and isotropic diffuse components, respectively, while for the modelling of background sources we employ the 4FGL-DR4 catalogue, adding a component for each source \citep[v34;][]{4FGL, Ballet23}. Energy dispersion correction is applied to all sources except the isotropic diffuse component. In this setup, the detection significance is evaluated considering the test statistic $\rm{TS} = - 2\;\ln L_{\textrm{max},0} /L_{\textrm{max},1}$, where $L_{\textrm{max},0}$ is the likelihood value for the null hypothesis and $L_{\textrm{max},1}$ the likelihood for the tested model. The TS follows a $\chi^2$ distribution, and thus $\rm{TS}=25$ approximately corresponds to a detection at $5\sigma$ and $4\sigma$ for 1 and 4 source parameters, respectively. 
    
\begin{figure}
  \includegraphics[width=0.95\columnwidth]{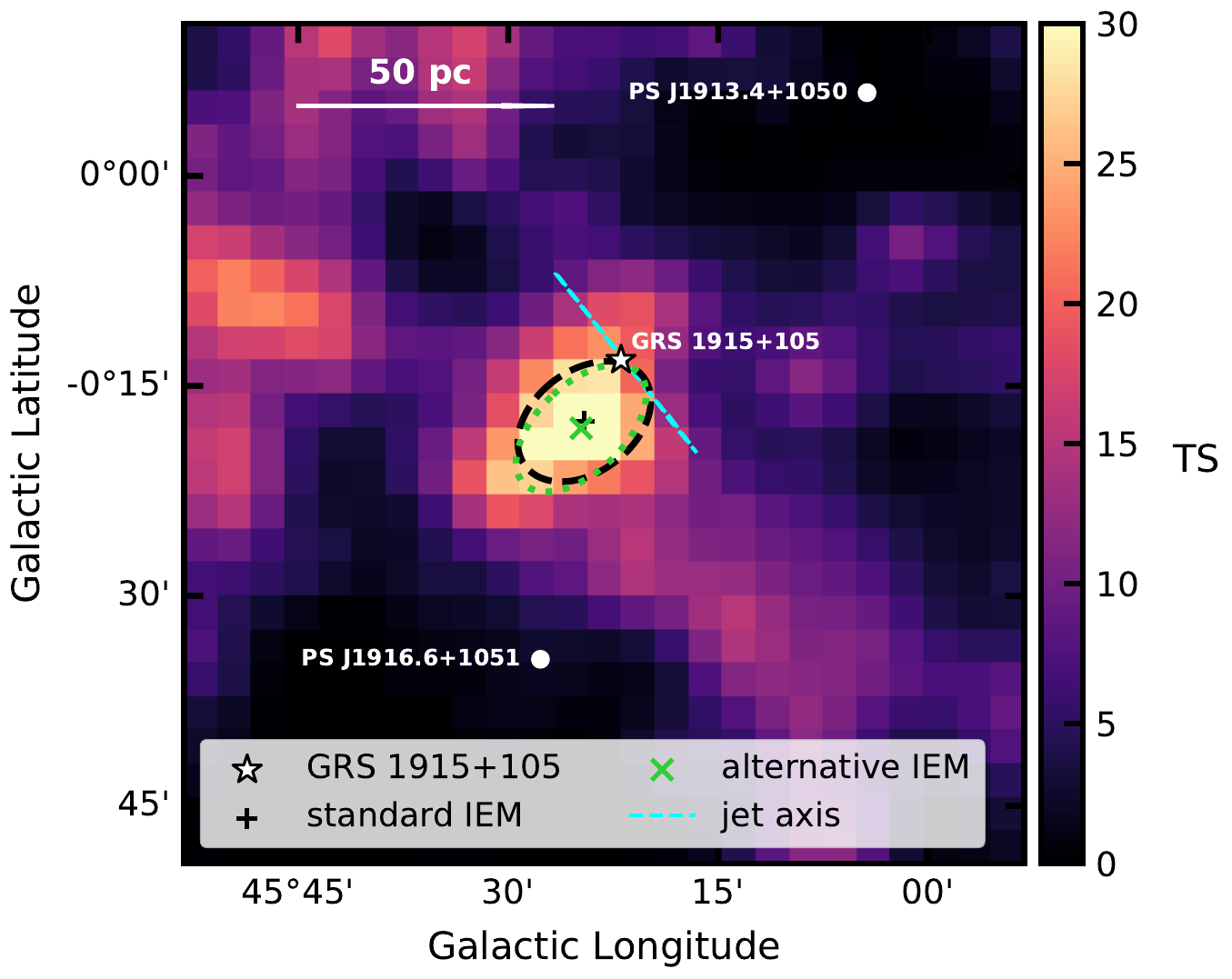}
  \caption{\label{fig:fig1}TS residual map (excluding \ours) of the ROI zoomed to the position of \grs, which is marked with a star. The best-fit position and 99\% statistical containment region of \ours are shown with a cross and dashed black line for the standard IEM and with a cross and dotted green line for the alternative IEM. The other two point-like excesses, which fall below the detection threshold when using the alternative IEM are indicated with white circles. The direction of the jet axis (e.g.~\cite{Reid2014}) is indicated with a straight dashed blue line.}
\end{figure}

Given the large source density in the region and the addition of substantial data not accounted for in the 4FGL-DR4 catalogue (almost two years), the likelihood fit is performed as follows: (1) we optimise the model of the ROI by fitting the normalisation of all sources in descending order of their predicted number of counts ($N_{\rm pred}$) down to $N_{\rm pred}=1$ (through the \texttt{fermipy.optimize} function). Then, (2) we free the normalisation of all model components within $5^{\circ}$ of the centre of the region, and all parameters for those sources within  $2.5^{\circ}$ or with $\rm{TS} > 500$, and fit them with \texttt{NEWMINUIT}. Finally, (3) we search for new sources ($>4\sigma$, $>0.2^{\circ}$ from known sources), and refit the ROI repeating step (2). These steps are performed twice to stabilise the fit. The procedure results in three new model components  close to \grs not present in the catalogue, required to explain three ($>4\sigma$) gamma-ray excesses in the region. In addition to the TS residual map, we also validate the goodness of our fit through a p-value statistic (PS) data/model deviation estimator \citep[Figure \ref{fig:SED}, B-C;][]{Bruel21}. The distribution of this estimator follows the expected distribution ($\propto 10^{-a\vert x \vert}$, for which we expected $a\sim 1$ and find $a=1.08$), which allows us to conclude that the region is well modelled. This does not mean that all the new model components not present in the catalogue are necessarily of physical origin. For example, they could arise from model defects of the Galactic interstellar emission model (IEM). To address this issue, we perform an additional analysis with gll\_iem\_v06.fits as an alternative IEM, the diffuse model employed in the latest catalogue of hard GeV sources \citep[3FHL,][]{3FHL}. Of the three excesses, shown in Figure~\ref{fig:fig1} and labelled \ours, PS J1916.6+1051 and PS J1913.4+1050, only the first is detected with TS $>25$ regardless of the IEM, suggesting \ours is the only excess of astrophysical origin.

\section{Results} \label{sec:results}

\begin{figure*}[h]
\includegraphics[width=\textwidth]{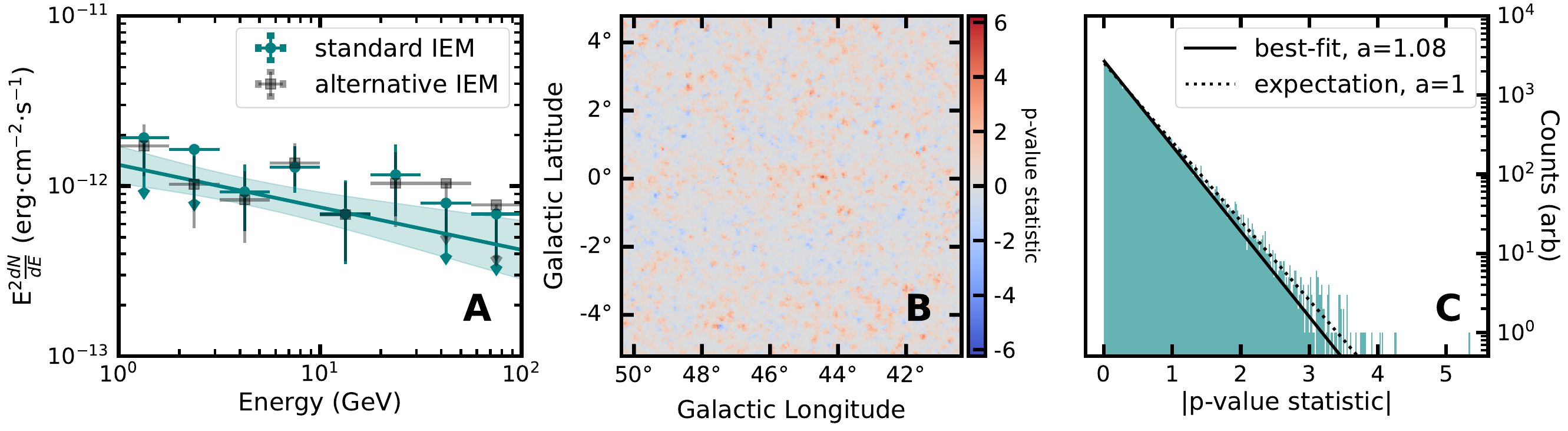}
\caption{\label{fig:SED} \textbf{A: }Best-fit spectral model and flux points for \ours. The teal circles show the flux points calculated with the standard IEM and the grey squares correspond to the alternative IEM. Uncertainties are statistical only. The machine-readable values for the flux points shown can be found in Table~\ref{tab:flux_points}. \textbf{B: }Residual p-value statistic map showing no significant excesses. For reference, PS$=6.24$ is equivalent to $5\sigma$. \textbf{C: }Distribution of the residual p-value statistic. The expected distribution is shown with a dotted line, the fitted distribution is shown with a solid line. }
\end{figure*}

Here we present a study on \ours, detected with TS $=33.6$ with the standard IEM and TS $=47.9$ when using the alternative one. It is best described by a point-like spatial model with best-fit position $l$, $b$ $= (45.409 \pm   0.026$,   $-0.292 \pm  0.024$). Interestingly, it is a potential counterpart for \grs, as the known microquasar lies at the edge of the source's containment uncertainty ($0.07^{\circ}$ at 99\% containment, statistical uncertainty only; Appendix \ref{appendix:morphological}). Regarding its spectral features, \ours is well described with a power law with $\Gamma= 2.25\pm 0.14$ (Figure \ref{fig:SED}, A), and an integrated photon flux above 1 GeV of $(6.74 \pm 0.33) \times 10^{-10}$ photons cm$^{-2}$ s$^{-1}$, equivalent to $(3.67 \pm 0.19) \times 10^{-12}$~erg~cm$^{-2}$~s$^{-1}$. We do not find significant evidence for curvature ($\Delta \rm{TS} = 2.6$) nor an exponential cutoff ($\Delta \rm{TS} = 3.4$) in the spectrum. The quoted uncertainties are statistical only - systematic uncertainties on the flux are mostly dependent on the IEM and of similar amplitude as the statistical uncertainties for standard LAT analyses \citep{Acero15}.

\subsection{Contamination from nearby sources}\label{subsec:the_snr}

The region surrounding \ours is densely populated by gamma-ray sources and possible counterparts other than \grs. We discuss associations with sources known via other wavelengths in Section \ref{subsec:counterparts}. Here we discuss instead the possibility that the new excess is due to poor modelling of one of the known nearby GeV sources. In particular, the brightest nearby ($<0.5^{\circ}$) source to \ours is 4FGL J1916.3+1108, reported to be the GeV counterpart of the supernova remnant (SNR) G045.7-00.4~\citep{thesnrpaper}. We carry out further morphological tests to determine if the GeV excess \ours can also be attributed to emission from the SNR when accounting for the spatial extension of the GeV emission from the SNR.

We assess the preference among several morphological models (each with $k$ degrees of freedom) by means of the Akaike information criterion \citep[AIC;][]{Akaike98}. This is defined as $\Delta$AIC$=2(\Delta k - \Delta \ln L)$. Although the improvement is relatively marginal, the best (i.e. smaller $\Delta$AIC) result is obtained by replacing the point-like spatial model for 4FGL J1916.3+1108 used in the 4FGL-DR4 catalogue with a disk, alongside a point-like component for \ours (Table \ref{tab:likelihood}). That is, the presence of the new source is preferred even when accounting for the extension of the neighbouring source of $r = 0.23^{+0.14}_{-0.07}$ ($\Delta\rm{TS}_{\rm{ext}}=15.14$). This angular size is consistent with the projected size of SNR G045.7-00.4, as well as the results from \cite{thesnrpaper}. If we further test the extension of \ours, we derive a limit of $0.14^{\circ}$ at $95\%$ confidence ($\Delta\rm{TS}_{\rm{ext}}=4.5$). This value increases to $0.22^{\circ}$ if instead we consider the 4FGL J1916.3+1108 as a point-like source. The preference for an independent source is additionally supported by an analysis restricted to energies between 5--500~GeV  (Appendix~\ref{appendix:morphological}). Consequently, we can confidently state that the new GeV signal is a distinct source, inconsistent in its spatial and spectral properties with other known nearby sources -- notably 4FGL J1916.3+1108 ($\Gamma = 3.00 \pm0.11$), even when considering its extension.

\begin{table*}
\caption{Likelihood and AIC values for different morphological analyses. The baseline scenario considers the model derived in Section~\ref{sec:analysis} excluding 4FGL~J1916.3+1108 and \ours. The spectrum of 4FGL~J1916.3+1108 is always described with a log-parabola shape as in the catalogue. The position of 4FGL~J1916.3+1108 is fixed to the catalog value in all steps but the second. The position of \ours is always fixed to the one obtained in Section~\ref{sec:analysis}. We note that, when \ours is not included in the model, the best-fit position of 4FGL~J1916.3+1108 is always consistent with the reported 4FGL-DR4 position ($<0.09^{\circ}$, 99\% confidence level) and does not shift in the direction of \ours regardless of the spatial model considered.}
\center
\begin{tabular}{l l c c }
\hline\hline
Step & Model & $\Delta k$ & $\Delta$AIC \\
 $\;$ &$\;$ & $\;$ \\
\hline 
0 & Baseline & --- & --- \\
1 & 4FGL J1916.3+1108 (point-like, 4FGL-DR4 position) & 3 & -129.9 \\
2 & 4FGL J1916.3+1108 (point-like, free position)  & 5  & -129.7 \\
3 & 4FGL J1916.3+1108 (extended) & 6 & -168.2 \\
4 & 4FGL J1916.3+1108 (point-like) + \ours (point-like) & 9 & -153.0 \\
5 & 4FGL J1916.3+1108 (extended) + \ours (point-like, fixed spectrum) & 8 & -170.1 \\
6 & 4FGL J1916.3+1108 (extended) + \ours (point-like, free spectrum) & 10 & -169.3 \\
\hline
\end{tabular}


\label{tab:likelihood}
\end{table*}

\subsection{Search for variability}\label{subsec:varibility}

We further explore the temporal properties of the new gamma-ray source, as these could carry additional information about the underlying particle acceleration processes. To this end, we search for periodical signals and flaring episodes, both blindly and specifically to \grs. We employ the likelihood formalism developed by \cite{Kerr19} and find no significant periodical modulation nor any gamma-ray flares (Appendix \ref{appendix:temporal}). We also find no significant differences between the emission before and after June 2018, when the system changed its X-ray state. This implies that \ours is a persistent gamma-ray source.

\section{Discussion} \label{sec:discussion}
\subsection{Identifying the counterpart} \label{subsec:counterparts}

\begin{figure}[h]
\includegraphics[width=0.95\columnwidth]{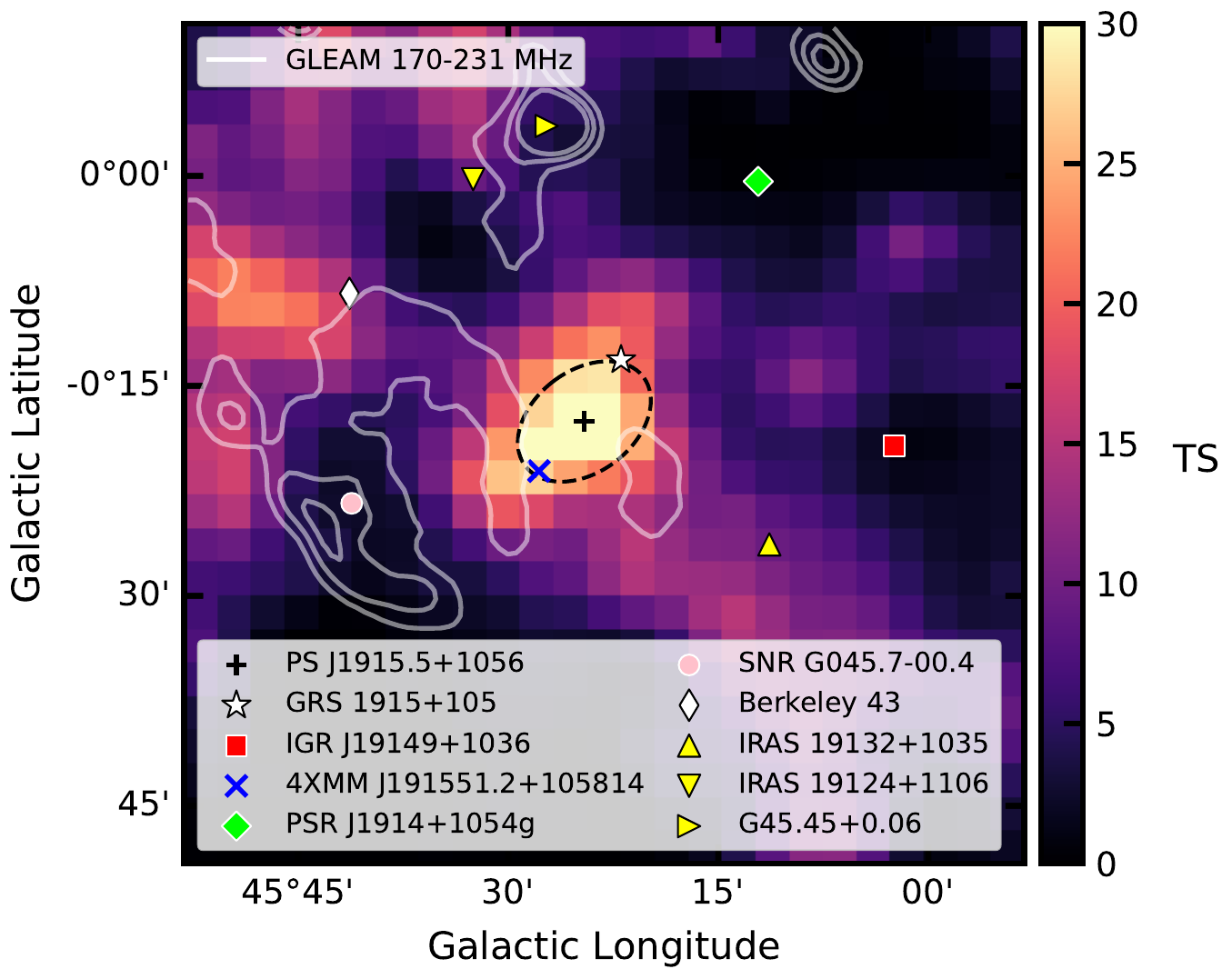}
\caption{TS map zoomed to the position of \ours with the location of possible counterparts.  A cross and dashed line represent the best-fit position and 99\% containment of \ours using the standard IEM. The position of \grs is represented with a white star. The position of the binary IGR J19149+1036 is shown with a red square. The white contours show radio emission from GLEAM~\citep{GLEAM} which outline the shape of a nearby SNR, indicated with a pink circle. The closest pulsar and stellar cluster are represented with a green rhombus and a white diamond, respectively. The only source besides \grs which falls within the 99\% containment radius is a faint unassociated \xray source, indicated with a blue x symbol. Finally, the three possible sites of jet interaction with the interstellar medium are indicated with yellow triangles. \label{fig:counterparts}}
\end{figure}

There are many classes of astrophysical objects which could be responsible for a point-like GeV excess in the Galactic plane. The most frequent association with point sources in the \fermi data are relatively powerful pulsars~\citep{FermiPulsars}. We make use of the ATNF Pulsar Catalogue~\citep{ATNF} and search for known pulsars in the vicinity of the excess. There are 4 pulsars inside a box of 1$^{\circ}$ width around the excess, three of which have too low spin-down power values ($\dot{E}<10^{33}$ erg~s$^{-1}$) to account for the excess~\citep{FermiPulsars}. 
The last object, PSR J1914+1054g, has $\dot{E}=4 \cdot 10^{35}$erg~s$^{-1}$~\citep{Motta2023} but lies more than 0.35$^{\circ}$ away from the best-fit position (see Figure~\ref{fig:counterparts}). Given the derived 99\% containment radius for the best-fit position is $0.07^{\circ}$, we can rule PSR J1914+1054g out as a likely counterpart.

SNRs can also be bright GeV sources. We query the SNRCat~\citep{SNRCat} catalogue and find that the only one within the 1$^{\circ}$ box is SNR G045.7-00.4, which lies 0.3$^{\circ}$ away from the best-fit position and is already accounted for by existing components in the model. The spectrum of the emission associated with the SNR has a softer spectrum ($\Gamma=3.00\pm0.11$) than the excess ($\Gamma=2.25\pm 0.14$), which allows to distinguish them (see Section~\ref{subsec:the_snr}). Within a SNR scenario, an asymmetric gamma-ray emission could be accounted for if particles escape from it and interact with nearby dense targets to produce the GeV excess. This has in fact been proposed for another source in the region, 4FGL J1914.5+1107c, whose  emission is correlated with dense CO gas at the distance of SNR G045.7-00.4~\citep{thesnrpaper}. However, the CO map at the distance of the SNR shown in Figure 6 of~\cite{thesnrpaper} reveals no dense target at the location of our GeV excess, ruling out this scenario. 

Additionally, several stellar systems within the region could act as particle accelerators. Young ($<10$ Myr) stellar clusters~\citep{Casse80, Ackermann11, VieuReville2023} are known to emit photons up to the TeV range, although the emission is usually extended. The closest known nearby cluster is Berkeley~43 ~\citep{HuntClusters}, which is not close enough to account for the excess, at more than 0.3$^{\circ}$ away from the best-fit position (see Figure~\ref{fig:counterparts}) with an angular extent of 5 arcminutes. Besides \grs, there is another known \xray binary system in the region of interest, IGR J19149+1036~\citep{labinaria}. The GeV excess exhibits no periodicity following the orbital period of IGR J19149+1036 of $22.25\pm0.05$ days (Figure~\ref{fig:period}). Furthermore, the binary is located more than 0.35$^{\circ}$ away from the best-fit position (see Figure~\ref{fig:counterparts}), so it can be safely discarded as a counterpart. 

While the region is filled with possible Galactic counterparts, we also investigate if the GeV signal could arise from an extragalactic source seen through the Galactic plane. There are no known AGN in the region in extragalactic catalogues such as NASA/IPAC or WISE~\citep{WiseCat}. However, this is most likely due to the fact that extragalactic catalogues and surveys usually exclude the Galactic Plane~\citep{NED_IPAC}, meaning that the sample of known AGN at low Galactic longitudes is highly biased to the brightest objects. We used the Fermi-LAT fourth catalogue of AGN ~\citep[4LAC;][]{4LAC} to estimate how many GeV-detectable AGN are expected in our region of interest. There are a total of 2863 objects in the 4LAC catalogue, which only considers high Galactic latitudes ($|b|>10^{\circ}$). This corresponds to a density of 275.7 objects per steradian, leading to 0.08 GeV-detectable AGN expected within the square of 1$^{\circ}$ width shown in Figure~\ref{fig:counterparts}. Taking into account the spectral properties of the observed excess, the expected number of extragalactic sources goes down by a factor of almost 2~\citep{4LAC}. We conclude that an extragalactic counterpart is unlikely but cannot at present be fully ruled out. 

Finally, the true counterpart could also be an unidentified source. We searched among \xray catalogues for high-energy sources other than \grs within the GeV 99\% containment region. There are none other than \grs in the SRG/ART-XC all-sky catalogue~\citep{Sazonov24,Pavlinsky21}, the INTEGRAL/IBIS 17-year catalogue~\citep{Krivonos21} and the second release of the \textit{Chandra} catalogue~\citep{Evans24}. In the second \textit{Swift} catalogue 2SXPS,~\citep[2SXPS,][]{Evans2020} there are two sources with good detection quality within the excess error radius - both with the same flux and both at the position of \grs, thus they both can be attributed to \grs. Finally, in the most recent XMM catalogue ~\citep[][DR14,]{XMM} there is one nearby source (besides \grs) that passes the good quality flag cut and has a detection likelihood greater than 10. The source is identified as 4XMM~J191551.2+105814 (R.A=19:15:51.30, Dec=+10:58:14.7) and it has a flux of $(1.25 \pm  0.19) \cdot 10^{-13}$ erg~s$^{-1}$~cm$^{-2}$ in the 0.2 to 12~keV energy range, orders of magnitude below that of \grs. No variability is reported for this source. The hardness ratio between the 2-4.5 and 4.5-12~keV bands is 0.1, indicating a relatively soft spectrum. The 
much fainter and softer \xray spectrum than \grs suggests a less powerful source and therefore a less likely counterpart -- although, again, this possibility cannot be completely excluded.

We conclude that \grs is the favoured counterpart for the new GeV source. In Section~\ref{subsec:power} we discuss a scenario where particles are accelerated in the jets of \grs relatively close to the central binary. Additionally, there have been suggestions that the \grs jets interact with the interstellar medium (ISM) far away from the central binary~\citep{Rodriguez98, Kaiser04, Tetarenko18}. There are three regions often invoked in this unconfirmed scenario: the two radio sources (IRAS~19132+1035 and IRAS 19124+1106) and the dense HII region G45.45+0.06 (all marked by triangles in Figure~\ref{fig:counterparts}). We do not detect gamma rays at the location of any of these sources.

\subsection{Power requirements} \label{subsec:power}
Gamma-ray emission is produced by relativistic particles interacting with their environment. We assume an acceleration region that produces a power-law spectrum of particles $dN/dE \propto E^{-2}\cdot exp({E/E_{\mathrm{cut}}})$ with a cutoff energy $E_{\mathrm{cut}}$ inside the jets of \grs and use the GAMERA code~\citep{Hahn2015, Hahn2022} to calculate the GeV emission in both a hadronic or leptonic scenario. It is unclear if \grs hosts an alternative source of power such an equatorial outflow \citep{Koljonen20}, and therefore we do not explore this scenario.

If protons are accelerated to relativistic energies in the jets, they would produce GeV radiation due to inelastic collisions with nearby gas. We use data from the FUGIN $^{12}$CO (J=1-0) survey based on observations at the Nobeyama Radio Observatory~\citep{FUGIN} to determine the density in the region. The details can be found in the Appendix~\ref{appendix:gas}. We find values ranging between $18 d_{9.4}^{-1}$ and $9 d_{9.4}^{-1}$~cm$^{-3}$ depending on the size of the region considered. For a value of the ambient density of $n\sim18$~cm$^{-3}$, relativistic protons need to reach energies of at least a few hundred GeV ($E_{\rm cut}\gtrsim250$ GeV) and have a total energy above 1~GeV of $\sim 3\cdot $10$^{49}$ erg to match the observed gamma-ray flux. If the average power supplied by the jets is 1\% of L$_{\mathrm{Edd}}$ and at least 10\% of that power (a total of$\sim10^{36}$~erg~s$^{-1}$) goes to acceleration of cosmic rays, supplying that energy would require $\sim1$~Myr. If the lower value of $n\sim9$~cm$^{-3}$ is considered, this time goes up by a factor two, as it is inversely proportional to the density. Given the estimated duration of the current binary evolution stage of \grs of 10.5~Myr~\citep{Belczynski2002}, this scenario should in principle be possible, although it is subject to great uncertainty given the unknown real duty cycle of the jets. Once accelerated, the particles would diffuse around the source - how far they can reach depends on value of the diffusion coefficient $D$. Assuming, as in~\cite{mitya}, an intermediate diffusion regime with $D\sim 10^{26}(E/10 \; \mathrm{GeV})^{0.5}$, 200~GeV protons would have reached scales of around 50~pc over one Myr. This corresponds to a 0.3$^{\circ}$ radius at the distance of \grs. Given the slight offset between \ours and \grs, it is plausible that particles escaped from \grs are causing an enhancement in the cosmic ray density around the source, which then translates to gamma-ray emission only where the target gas is dense enough~\citep{Aharonian2004,Valenti2005, mitya}.

If instead we assume the emission to be due to electrons, any consideration is subject to the high uncertainty on the magnetic field and radiation fields in the region. Modelling of the radio properties of \grs suggests a high magnetic field on small scales ($<1$~pc from the engine), on the order of 1~G, which would heavily suppress inverse Compton scattering from electrons~\citep{mitya}. However, the \fermi results do not tightly constrain the position within the jet. Considering the extension upper limit derived in Section~\ref{sec:results}, which corresponds to a radius $<36$~pc at the distance of \grs, the gamma-ray emission could arise from a different region where the magnetic field is considerably lower. Assuming a more favourable value of $B\sim 5\mu$G on the order of the ISM magnetic field~\citep{JanssonFarrar}, the interstellar radiation field from~\cite{Popescu2017} at the position of \grs and again an average jet power of 1\% of the Eddington luminosity but now with only 1\% of this power going to relativistic ($>1$~MeV) electrons, an injection time several orders of magnitude longer than the estimated age of~\grs of 10.5~Myr~\citep{Belczynski2002} would be needed to match the observed gamma-ray flux. One could instead consider an episode in which the 
accretion rate is higher, which translates to more power carried by the jet and thus available to electrons. Even in such a case, where the jet sustains powers of 10-100\% of L$_{\mathrm{Edd}}$, electrons would require 0.1--1\% of that power for at least 10~Myr to match both the observed flux level and spectral index. The prospects for leptonic emission improve if one considers additional sources of radiation such as the companion star or the accretion disk, with the caveat that these will dominate in regions close to the central engine, where the magnetic field might be much higher as well. Note that the combination of assumed magnetic field of $B\sim 5\mu$G and interstellar radiation field result in faster loss timescales for inverse Compton than synchrotron for the relevant energies. Higher values of $B\sim 10-20\mu$G such as those estimated at multi-pc scales from other microquasars~\citep{SafiHarb2022, HESSCollaboration2024} would drive up synchrotron losses and make it even more challenging to explain the GeV observations as inverse Compton emission.

\section{Conclusion} \label{sec:conclusion}
We have presented here evidence for a GeV counterpart (\ours) to the microquasar \grs. We find a point-like source with a relatively hard spectrum, for which we do not find significant transient or periodic flux modulations. The position of \ours is inconsistent with other known potential particle accelerators in the region, and its spectral and morphological properties are sufficiently different from those of the GeV counterpart to the nearby SNR G045.7-00.4 to exclude a common origin. Due to lack of dense gas at the distance of the SNR, we can dismiss the possibility that \ours is the result of protons that escaped the SNR. Instead we find that surrounding \grs there is enough gas to explain the observed emission as arising from the interaction of protons accelerated in the jets with the nearby gas. This scenario relies on a moderate duty cycle of the jet power over a relatively large timescale, which is largely unconstrained and subject to great uncertainty. A leptonic scenario is also plausible but disfavoured, as it requires a substantial amount of the jet power, even when assuming a low (5$\mu$G) magnetic field and the interstellar radiation field as targets for interaction. While we cannot fully discard the association with the much fainter \xray source 4XMM~J191551.2+105814 or an unknown blazar seen through the Galactic plane, we note that there is no evidence for a cutoff in the GeV spectrum of \ours, which suggests that the emission might extend to even higher gamma-ray energies. A detection in the multi-TeV band would rule out an extragalactic origin and further confirm the association with \grs, particularly given the improved angular resolution of Cherenkov telescopes. The firm identification of this microquasar as a gamma-ray emitter would establish LMXBs as high-energy particle accelerators and constrain their contribution to the cosmic ray content of our Galaxy.

\section{Acknowledgements}
\begin{acknowledgments} We thank Quentin Remy for his help with the gas density estimation, as well as Jian Li, Teddy Cheung, Matthew Kerr, and Philippe Bruel for their comments on the manuscript. This research has made use of the NASA/IPAC Extragalactic Database, which is funded by the National Aeronautics and Space Administration and operated by the California Institute of Technology.
This research has made use of the VizieR catalogue access tool, CDS,
Strasbourg, France \citep{10.26093/cds/vizier}. The original description 
of the VizieR service was published in \citet{vizier2000}.
The CO data were retrieved from the JVO portal (\url{http://jvo.nao.ac.jp/portal/}) operated by ADC/NAOJ. The Fermi LAT Collaboration acknowledges generous ongoing support from a number of agencies and institutes that have supported both the development and the operation of the LAT as well as scientific data analysis. These include the National Aeronautics and Space Administration and the Department of Energy in the United States, the Commissariat \`{a} l'Energie Atomique and the Centre National de la Recherche Scientifique / Institut National de Physique Nucl\'{e}aire et de Physique des Particules in France, the Agenzia Spaziale Italiana and the Istituto Nazionale di Fisica Nucleare in Italy, the Ministry of Education, Culture, Sports, Science and Technology (MEXT), High Energy Accelerator Research Organization (KEK) and Japan Aerospace Exploration Agency (JAXA) in Japan, and the K. A. Wallenberg Foundation, the Swedish Research Council and the Swedish National Space Board in Sweden. Additional support for science analysis during the operations phase from the following agencies is also gratefully acknowledged: the Istituto Nazionale di Astrofisica in Italy and the Centre National d'Etudes Spatiales in France. This work performed in part under DOE Contract DE-AC02-76SF00515.

\end{acknowledgments}

%

\vspace{5mm}
\facilities{\textit{Fermi}-LAT, Nobeyama Radio Observatory~\citep{FUGIN}.}


\software{astropy~\citep{astropy},  
          fermipy~\citep{Wood17}, 
          godot~\citep{Kerr19},
          gammapy~1.2~\citep{Donath2023, gammapy1p2},
          numpy~\citep{numpy},
          matplotlib~\citep{matplotlib}.
          }

\appendix

\section{Analysis details}\label{appendix:morphological}
The statistical 99\% containment region for the best-fit position when using the standard IEM in the analysis above 1~GeV (see Figure~\ref{fig:fig1}) is parameterised as an ellipse with semi-major and semi-minor axes of 0.088\degree~and 0.060\degree~respectively, rotated by an angle 143\degree~with respect to the Galactic North (eastward). For the alternative IEM the values are 0.095\degree, 0.052\degree~and 136\degree, respectively. A machine-readable version of the flux points shown in Figure~\ref{fig:SED} can be found in Table~\ref{tab:flux_points}.

\begin{table*}[h]
\caption{Flux point values for both the standard and alternative diffuse models}  
\center
\begin{tabular}{ccccccc }
\hline\hline
Energy & Energy range & Flux  & TS  & Flux  & TS\\
& & standard &standard &alternative& alternative \\
 GeV & GeV & 10$^{-12}$ erg cm$^{-2}$ s$^{-1}$& & 10$^{-12}$ erg cm$^{-2}$ s$^{-1}$ &\\
\hline 
1.33 & 1-1.78 & $<1.922$  & 1.99 & 1.716 $\pm$ 0.587 &  8.80\\
2.37 & 1.78-3.16 & $<1.642$  & 2.63& 1.022 $\pm$ 0.466 & 4.92\\
4.22 & 3.16-5.62 & 0.925  $\pm$ 0.401 & 6.17 & 0.830 $\pm$ 0.381&  5.38\\
7.50 & 5.62-10 &  1.289 $\pm$ 0.399  &  15.42& 1.367 $\pm$ 0.392& 18.81\\
13.33 & 10-17.78 &  0.686 $\pm$  0.637&  5.74& 0.680 $\pm$ 0.344& 6.41\\
23.73 &17.78-31.62 & 1.165 $\pm$ 0.547 & 9.09&  1.036 $\pm$ 0.505 & 8.61\\
42.17 & 31.62-56.23& $<0.794$  &0 & $<1.038$  & 0.39\\
74.99 & 56.23-100 & $<0.685$  & 0 & $<0.777$ & 0 \\
\hline
\end{tabular}
\label{tab:flux_points}
\end{table*}

In order to further verify that \ours has different properties than the emission from the nearby SNR, we repeat the analysis with a low energy threshold of 5~GeV. At those energies, \ours is still detected at TS = 31, while the TS from 4FGL J1916.3+1108 goes down from TS $ = 139$ to TS $= 16$. When fitting the region without the additional model component for \ours, the new best-fit position for 4FGL J1916.3+1108 is found to be at ($l$,$b$) = ($45.432 \pm   0.028$, $-0.298 \pm  0.028$), and thus consistent only with the position of \ours (Appendix \ref{appendix:morphological}). 

\begin{figure}[h]
\includegraphics[width=0.32\textwidth]{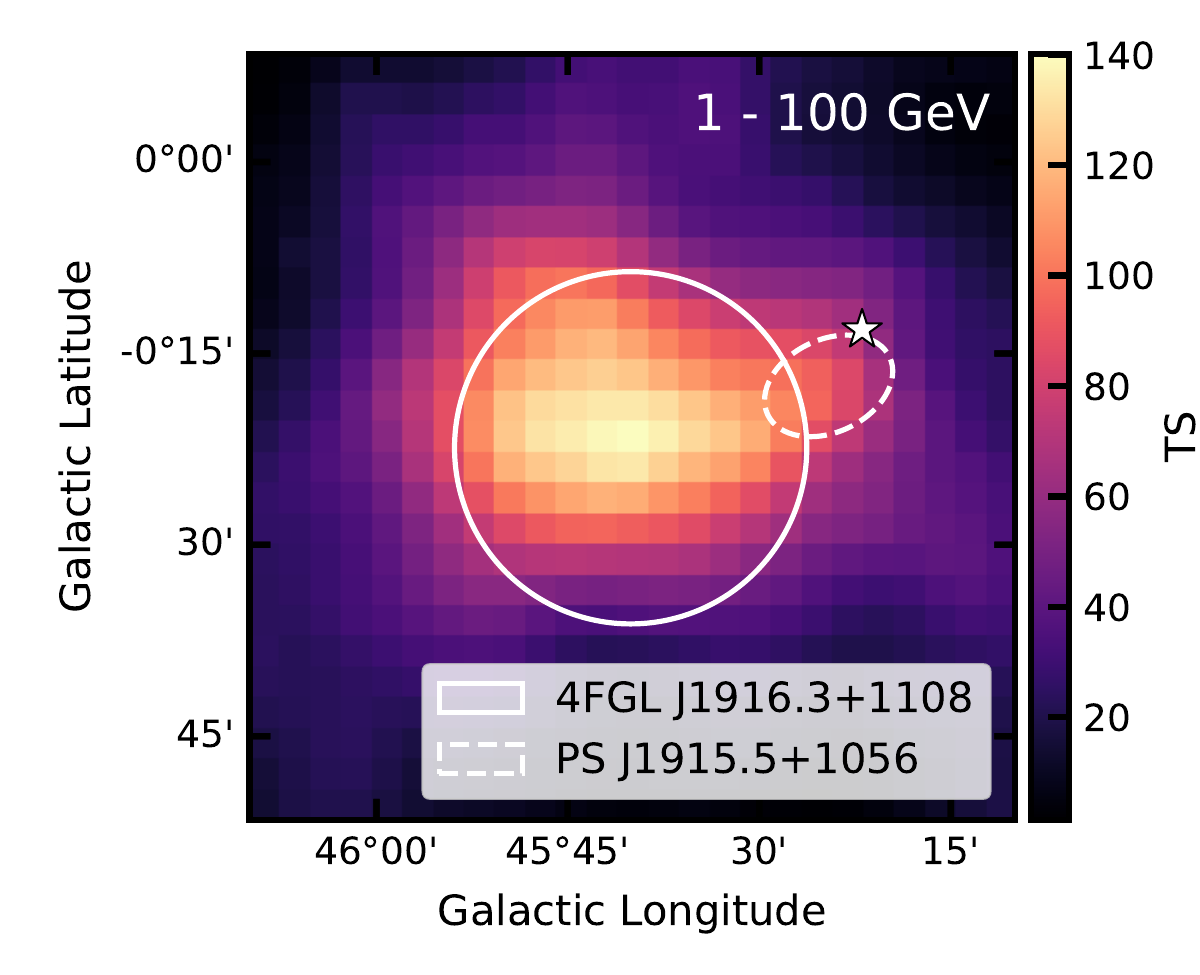}
\includegraphics[width=0.32\textwidth]{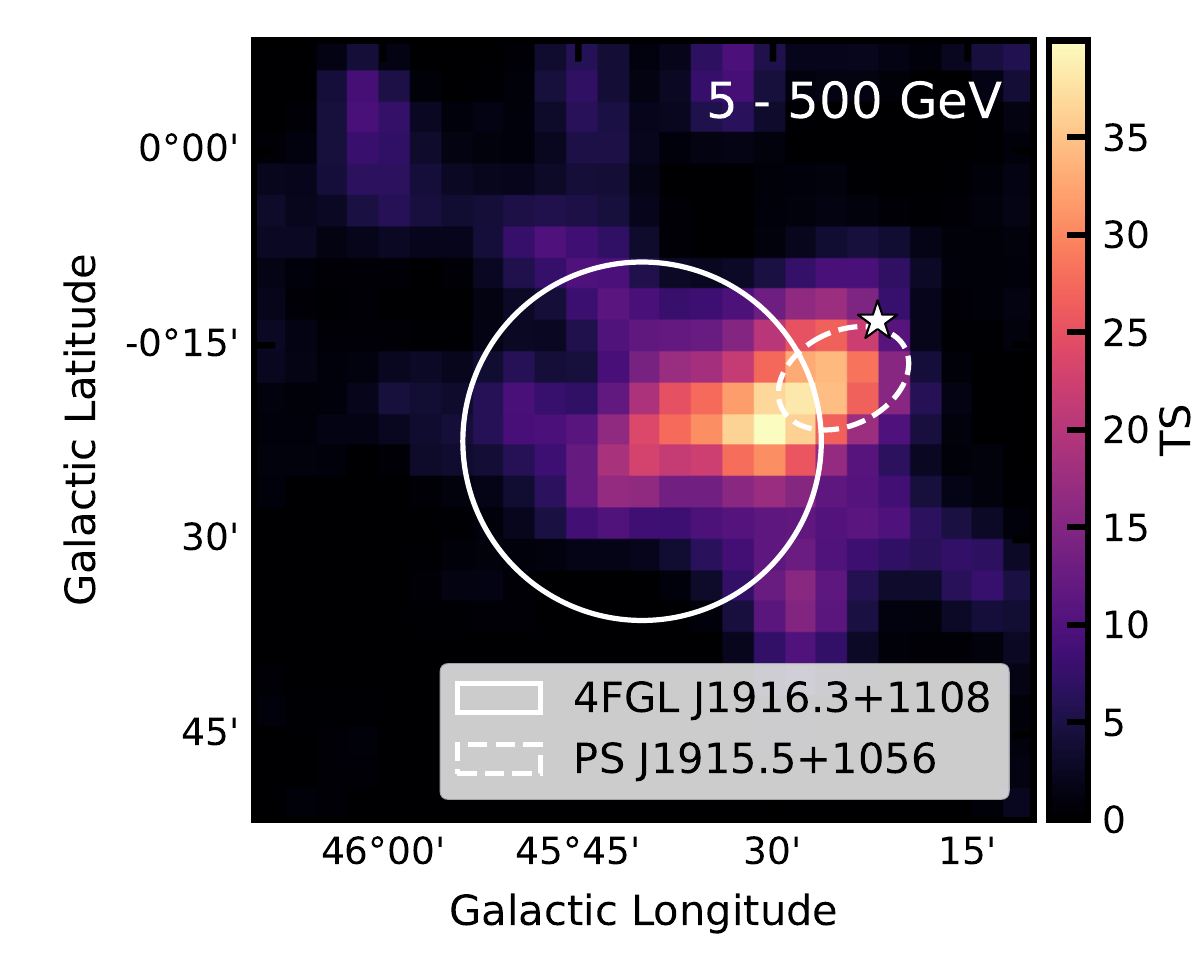}
\includegraphics[width=0.32\textwidth]{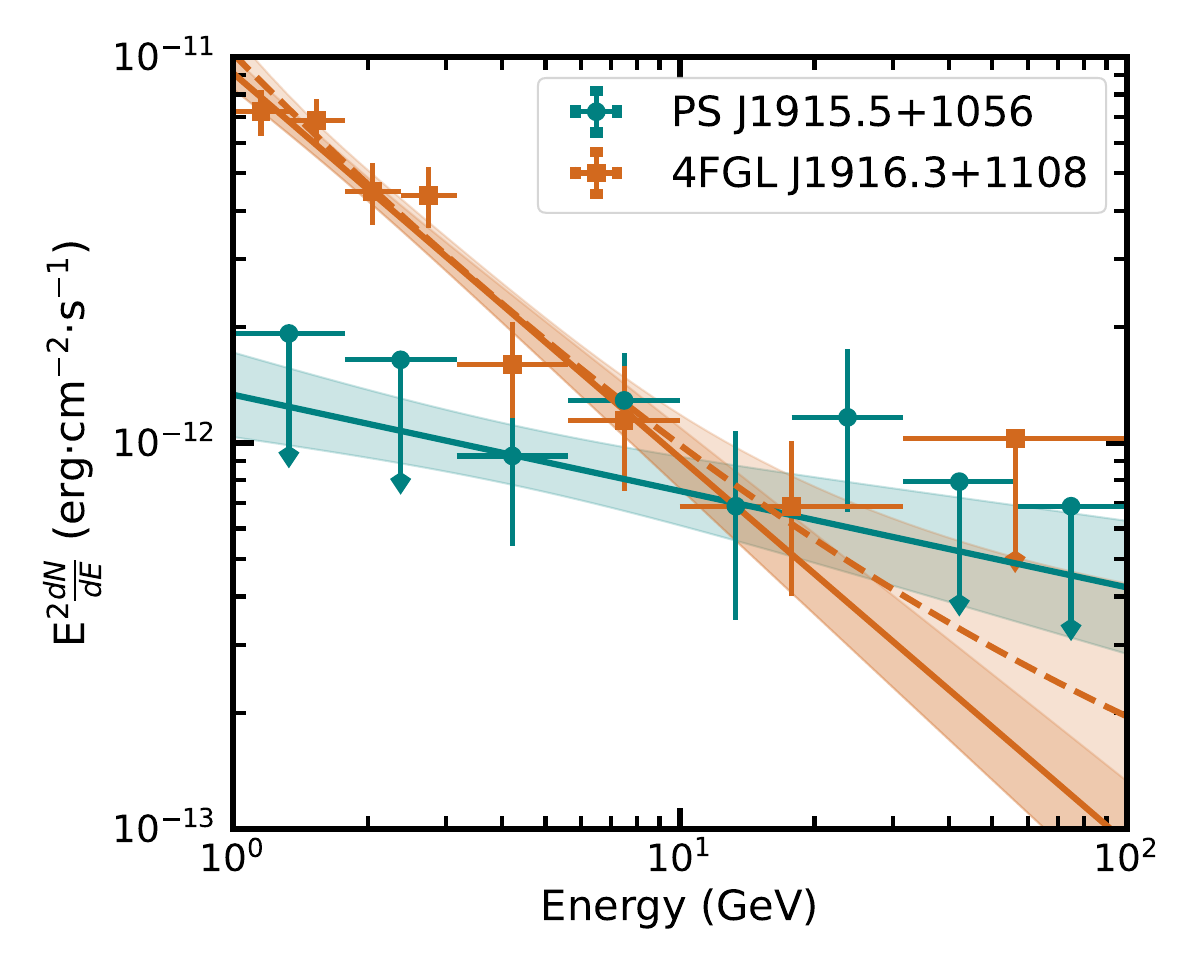}

\caption{The first two panels show TS residual maps after considering all the model components except those associated with \ours and SNR G045.7-00.4, centred at the catalogue position of 4FGL J1916.3+1108. Comparison between the  1 -- 100 GeV (\textit{left}) and 5~--~500~GeV (\textit{middle}) energy ranges reveal energy dependence of the overall morphology of the gamma-ray signal of the region. The solid line represents the best fit extension of 4FGL J1916.3+1108, while the dashed line represents the 99\% position uncertainty of \ours as obtained in the analysis above 1 GeV (Section \ref{sec:results}). The white star represents the nominal position of \grs.\label{fig:morphology}. The remaining panel (right) shows the best-fit spectral model and flux points for the 4FGL J1916.3+1108 power law (solid line) and log-parabola (dashed line) model. The SED of \ours (standard IEM case) is also shown for easy comparison.}
\end{figure}

\section{Search for variability with photon weights}\label{appendix:temporal}

\begin{figure*}[h]
\includegraphics[width=\textwidth]{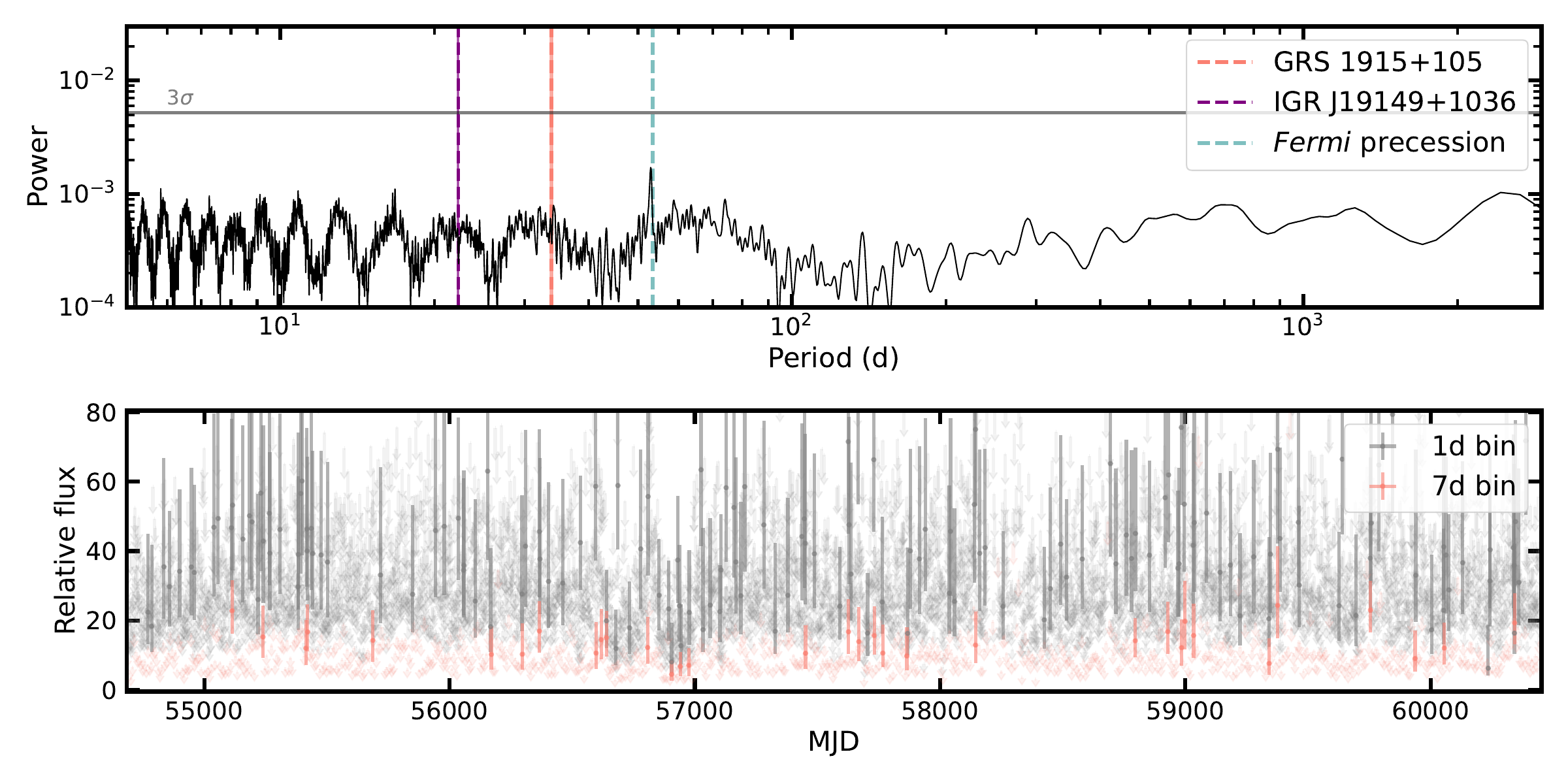}
\caption{Temporal analysis of PS J1915.5+1056. (\textit{Top}) Periodogram for PS J1915.5+1056, derived using photon weights. The orbital periods of \grs and IGR~J19149+1036 are marked (with the uncertainty represented with a shaded area), as well as \textit{Fermi}'s orbital precession period. Globally, no signal is found above 3$\sigma$ for any frequency evaluated. (\textit{Bottom}) Light curve for PS J1915.5+1056 employing a daily and weekly binning. Flux values are displayed relative to that obtained in Section~\ref{sec:results}. We report detections for bins with $\rm{TS}>4$, otherwise upper limits are reported at 95\% confidence level. The results are consistent with a constant flux.
\label{fig:period}}
\end{figure*}

We carry out a blind search for state changes by incorporating photon weights into the Bayesian Block algorithm \citep{Scargle13, Kerr19}. Thus, we apply it to the underlying likelihood distribution itself and not to the derived flux points. We select all photons above 100 MeV within $1^{\circ}$ of the best-fit position of the source, extending the maximum zenith angle to $105^{\circ}$. The previously derived model ROI is then employed to compute the probability for each photon to be associated with the new source, which is then used as a weight to compute the number of distinct blocks in our light curve ($N_b$). We penalise the addition of further blocks by means of a prior $\propto N_b^{\gamma}$. Consequently, the parameter ${\gamma}$ sets the significance of flux changes. We derive the false positive rate by redistributing photons randomly among cells \citep{Kerr19}. In order to ensure zero false detections within the 16 years of data (so, a false alarm rate of $10^{-4}$), the prior has to be set to $\gamma=10$. From this, we find the data to be well represented by a single cell, and therefore the source does not flare significantly (Figure \ref{fig:period}). This is consistent with the results by \cite{Bodaghee2013} using 4 years of data. Additionally, we divide the dataset into two ranges: before and after the state change in \grs that occurred in July 1st 2018~\citep{Motta2021, Balakrishnan2021}. We perform a regular likelihood analysis on each of these ranges, finding no significant distinction between them. Before the state change, the excess has a significance of $\rm{TS}=23.0$ and a flux above 1~GeV of $(3.83 \pm 0.94) \times 10^{-12}$ erg cm$^{-2}$ s$^{-1}$. After the state change the values are $\rm{TS}=11.4$ and $(3.41 \pm 1.19) \times 10^{-12}$ erg cm$^{-2}$ s$^{-1}$ for significance and flux, respectively.

We search for variability that follows the orbital period of \grs by phase-folding the original dataset with said period of \citep[$P=33.85$ d, $T_0=55458.68$ MJD; ][]{Steeghs2013}. We consider 2 and 4 bins of orbital phase and re-fit the model of the ROI with the standard likelihood analysis above 1 GeV as described in Section \ref{sec:analysis}. We find no flux modulation along the orbit. Finally, we employ the previously derived weights to carry out a blind search for periodical modulations by means of a weighted aperture photometry light curve \citep{FermiLAT12}. We derive the Lomb-Scargle periodogram, shown in Figure~\ref{fig:period}, which shows no signal in the power distribution above $3\sigma$. The largest peak is found at the 53.7 d precession period of \textit{Fermi}. No significant period is found either when using the likelihood formalism from \cite{Kerr19} which considers background periodical fluctuations as well.

\section{Estimation of the local density}\label{appendix:gas}
\begin{figure*}[h]
\includegraphics[width=\textwidth]{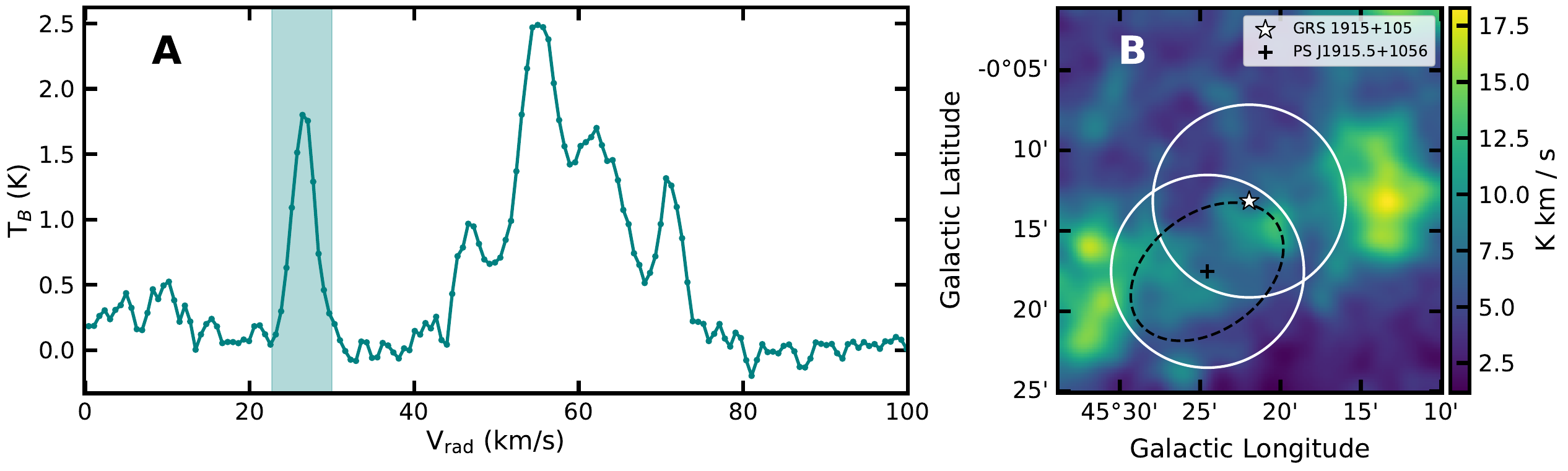}
\caption{Molecular gas emission around \grs. \textbf{A:} FUGIN $^{12}$CO (J=1-0) spectrum of molecular gas towards \grs. \textbf{B:} Integrated $^{12}$CO emission map in the velocity interval shaded green in panel A. The white circles represent the extraction regions of 0.1\degree~radius used to derive the number density. The dashed black ellipse and cross show the position and 99\% position uncertainty of \ours. The white star represents the nominal position of \grs.
\label{fig:gas}}
\end{figure*}

We use data from the FUGIN $^{12}$CO (J=1-0) survey based on observations at the Nobeyama Radio Observatory~\citep{FUGIN} to determine the density in the region. We extract the spectra of $^{12}$CO in a box with side 0.4\degree~centred on \grs. The resulting spectrum is shown in Figure~\ref{fig:gas} and shows a peak at V$_{\rm LSR}=26.5$~km~s$^{-1}$, which corresponds to a near distance of $1.6\pm 0.4$~kpc and a far distance of $10\pm 0.4$~kpc~\citep{Brand1993,Reid2019,Wenger18,Wenger2021}, the latter consistent with the distance to \grs of $d =9.4\pm0.6\pm0.8$~kpc~\citep{Reid2023}. Note that this peak remains even if the spatial extraction region for the spectra is reduced to a box of side 0.1\degree. We integrate this peak in each pixel of the box, between velocities of 22.7 and 30~km~s$^{-1}$, which results in the map shown in Figure~\ref{fig:gas}. Adopting a CO-to-H$_2$ conversion factor X$_{CO}$ following the functional form in~\cite{CO-conversion} and using the same parameters as in~\cite{ctagps}, results in a conversion factor X$_{CO} = 8.45\cdot10^{19}$ s~K$^{-1}$~km$^{-1}$~cm$^{-2}$ at the distance of \grs. Following the procedure outlined in e.g.~\citet{gas_procedure}, we derive a total mass inside of a region of 0.1$^{\circ}$ radius around the best-fit position of the new GeV source of $1.1\cdot 10^5 d_{9.4}^2 M_{\odot}$ (within the \fermi PSF at 10~GeV). Assuming that the mass distribution is homogeneous within a sphere, this corresponds to a hydrogen number density of n$=19d_{9.4}^{-1}$~cm$^{-3}$. To explore how robust this estimate is, we select a region of the same size around \grs, for which the resulting density is n$=17d_{9.4}^{-1}$~cm$^{-3}$. If instead we consider a larger radius corresponding to the extension upper limit of 0.22$^{\circ}$ (Section~\ref{sec:results}), the resulting density is lower by a factor 2. 

\bibliography{grs1915.bib}{}
\bibliographystyle{aasjournal}
\end{document}